\newcommand{\gaga}{$\gamma\gamma$ \hspace*{0.1mm}}
\newcommand{\jpsi}{J/$\psi$ \hspace*{0.1mm}}
\newcommand{\rrho}{$\rho^0$ \hspace*{0.1mm}}
\documentclass{PoS}
\usepackage{lineno}

\title{Measurements of vector meson photoproduction with ALICE in ultra-peripheral Pb-Pb collisions at $\bf{\sqrt{s_{\rm NN}} = 5.02}$ TeV}

\ShortTitle{Vector meson photoproduction with ALICE}

\author{\speaker{V. Pozdniakov} \\
        On behalf of the ALICE Collaboration \\
        JINR (Dubna, Russia)\\
        E-mail: \email{Valeri.Pozdnyakov@cern.ch}}

\author{Yu. Vertogradova\\
        JINR (Dubna, Russia)\\
        E-mail: \email{Iuliia.Vertogradova@cern.ch}}

\abstract{The intense photon fluxes of relativistic nuclei provide a possibility
to study photonuclear and two-photon interactions in ultra-peripheral collisions (UPC)
where the nuclei do not overlap and no strong nuclear interactions occur. \\
The study of such collisions provides information about the initial state of nuclei. 
First ALICE results from the LHC Run 2 are presented for forward
exclusive \jpsi production in UPC, which is sensitive to the gluon distribution in nuclei.
The increased statistics and the higher collision energy allows for a more detailed study at
lower values of Bjorken-$x$. \\
The analysis of the $\gamma+A \rightarrow \rho^0+A$ process in UPC is a tool to test the so-called black disk regime,
where the target nucleus appears like a black disk and the total $\rho^0+A$ cross section reaches its limit.
ALICE reports new measurements of \rrho photoproduction cross sections in Pb-Pb UPC at $\sqrt{s_{\rm NN}}=5.02$ TeV
at mid-rapidity, which are compared to predictions.}

\FullConference{EPS-HEP 2017, European Physical Society conference on High Energy Physics\\
		5-12 July 2017\\
		Venice, Italy}

\begin{document}

\section{Introduction}
\par
\begin{figure}[!h]
\begin{center}
\vfill \begin{minipage}{0.57\linewidth} \vspace*{-0.0cm} \hspace*{-0.5cm}
\mbox{\includegraphics[width=1.1\linewidth]{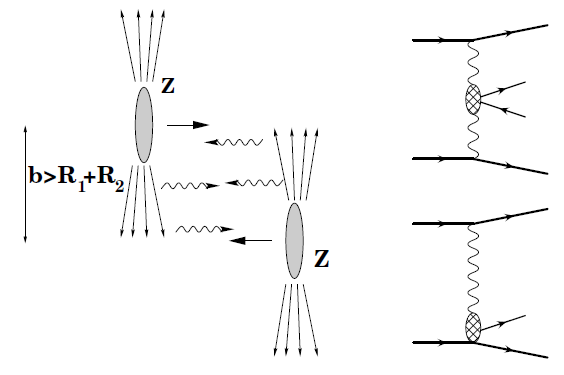}} \vspace*{-0.0cm}
\end{minipage} 
\vspace*{-0.7cm}
\caption{\label{diag} Ultra-Peripheral Collisions (UPC) of heavy ions (left) and diagrams of
two-photon (right top) and photonuclear (right bottom) UPC.}
\end{center}
\end{figure}

 Interactions of relativistic colliding heavy ions are a rich and fruitful field for investigations.
One of the main parameters of the collisions, which determines the physics nature of interactions, is the impact parameter.
If the impact parameter is larger than the sum of the nuclear radii, the so-called Ultra-Peripheral Collisions (UPC) can occur (Fig.~\ref{diag}).
In this case the nucleon electric charges work coherently and produce intense and (in the case of the Large Hadron Collider, LHC) energetic
photon fluxes which can be described by an equivalent photon approximation (EPA) \cite{bud}.
\par The experimental conditions in heavy-ion collisions generate two important characteristics of UPC. The coherent behavior of protons
in a nucleus makes the photon fluxes stronger by a factor of $Z^2$ compared to proton beams. Also
the coherence restricts the photon virtualities to very small values ($Q^2 = -q^2 \leq 1/R^2$, where $R$ is the radius of the nucleus)
due to their strong suppression by the nuclear electromagnetic form factor and almost zero scattering angles of ions.
Thus the photons can be considered as quasi-real.
\par The photon induced reactions at the LHC can be presented either due to pure electromagnetic photon-photon processes
or due to photon-nuclear reactions (diagrams in Fig.~\ref{diag}, right). The latter corresponds to the process when the photon fluctuates
to a bound $q\overline{q}$ system (a vector meson) which then elastically scatters (via Pomeron exchange) off the nucleus.
The total cross section can be factorized into the photon flux and the cross section of the two-photon or photonuclear reaction.
\par The study of UPC requires high-energy beams (extending the photon flux to high energies),
large pseudorapidity coverage of the experimental setup,
and a special trigger configuration. UPC results obtained so far by RHIC and LHC experiments \cite{prev} 
are limited and new experimental data are needed to extend UPC studies to new domains.
\par This paper presents ALICE results on \jpsi and \rrho photoproduction cross section measurements 
in Pb-Pb UPC at $\sqrt{s_{\rm NN}}=5.02$ TeV.

\section{Experimental setup}

A detailed description of ALICE is given elsewhere \cite{alice}. ALICE is designed to measure particles over a
wide kinematic range. Only the sub-detectors relevant to vector meson photoproduction measurements are shortly described below:
\begin{itemize}
\item Muon Spectrometer (MS) to reconstruct large rapidity (-4 $< \eta <$ -2.4) muons from \jpsi decays. It consists 
of a composite (layers of both high- and low-Z materials) absorber starting 90 cm from the vertex, a large dipole
magnet with a 3 Tm field integral and ten planes of high-granularity cathode strip tracking stations. 
A second muon filter at the end of MS and four planes of RPC are used for muon identification and triggering;
\item Inner Tracking System (ITS), a six-layer, silicon vertex detector, and the Time-Projection Chamber (TPC) 
to measure final state particles from \rrho decays at mid-rapidity. The TPC provides d$E$/d$x$ resolution of better
than 5-7\% and the TPC can serve, in addition to tracking, as a detector for particle identification;
\item trigger detectors - Silicon Pixel Detector (SPD) of ITS, forward scintillator detectors and trigger chambers of MS. 
The hardware trigger in ALICE combines the input from detectors with fast-trigger capabilities and it operates at several levels to
satisfy the individual timing requirements of the different detectors.

\end{itemize}

\section{Coherent \jpsi photoproduction}

\par Photoproduction of \jpsi on a proton in $\gamma$ + p $\rightarrow$ \jpsi + p reactions is well modeled 
in perturbative QCD by the exchange of two gluons \cite{photonp} and experimental data obtained by HERA experiments were used
to constrain the gluon PDF in the proton at low Bjorken-$x$ \cite{hera}. Exclusive production of a heavy vector meson, 
like the J/$\psi$ and the $\Upsilon$, in
heavy-ion interactions probes the nuclear PDF of gluons \cite{npdf} which has considerable uncertainty at low-$x$.
At the forward \jpsi rapidities studied here (-4 $< y <$ -2.5), the relevant values of $x$ are $\simeq 10^{-2}$ 
and $\simeq 10^{-5}$. The two values reflect the fact that either nucleus can serve as photon emitter or photon target.
\par The analysis is based on a sample of events collected during the 2015 Pb-Pb LHC run,
selected with a dedicated trigger (CMUP). The integrated luminosity corresponds to about 216 $\mu$b$^{-1}$.
The CMUP requires two muons with transverse momentum p$_{\rm T}$ above 1 GeV/{\it c} and no hits in the forward scintillators to reject hadronic collisions.
The selection of the muons is based on a set of detector specific criteria of muon reconstruction quality including the muon track matching to the trigger.
An event candidate to the coherent \jpsi photoproduction is then selected by physically motivated criteria: the event has to 
contain a pair of muons (dimuon) with opposite electric charges and p$_{\rm T} <$ 250 MeV/{\it c} to remove the majority 
of the incoherently produced \jpsi events.
\par
\begin{figure}[!h]
\vfill \begin{minipage}{0.4\linewidth} \begin{center} \vspace*{0.cm} \hspace*{-1.cm}
\mbox{\includegraphics[width=1.1\linewidth]{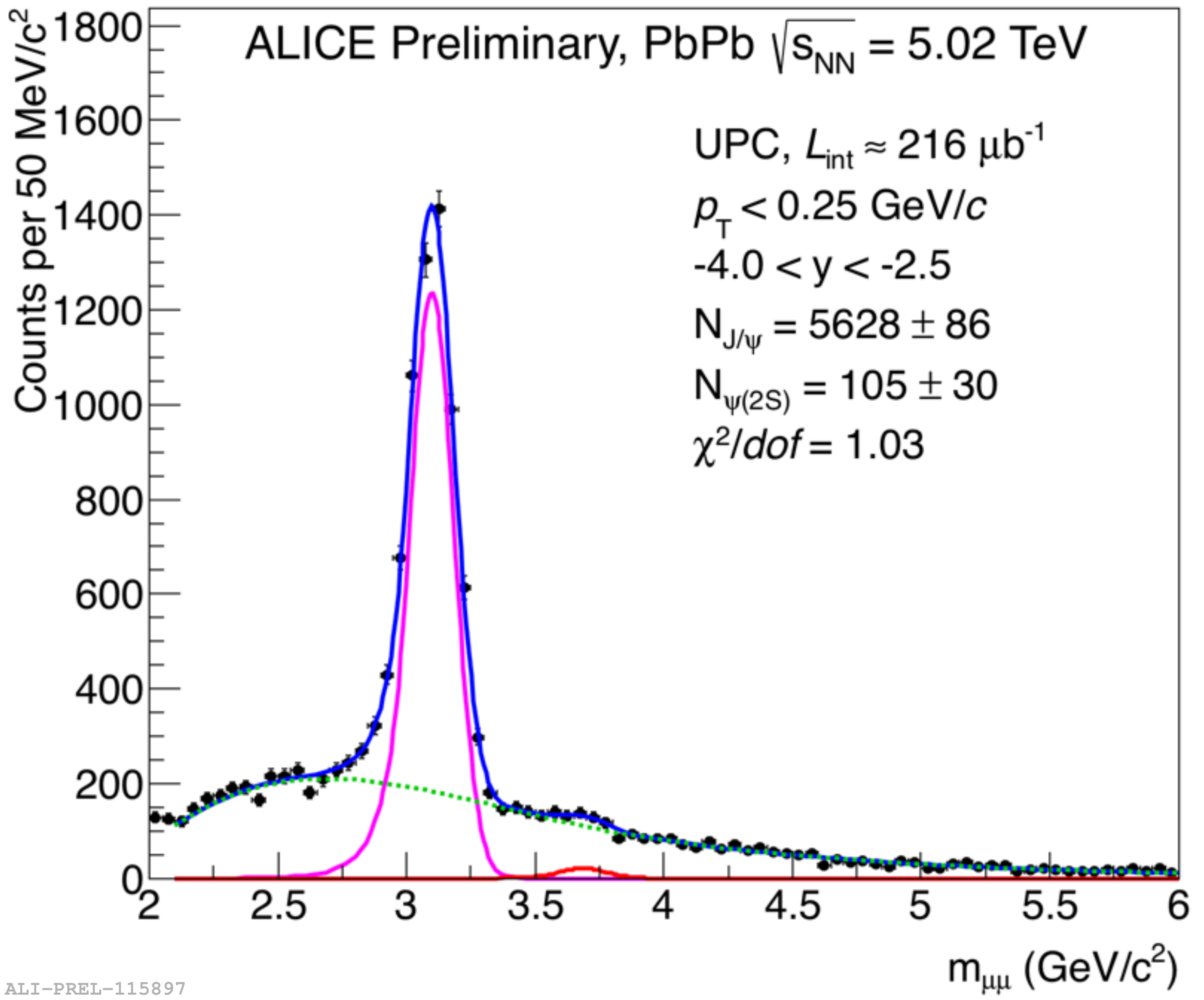}} \end{center} \vspace*{-0.cm}
\caption{\label{jpsisignal} Invariant mass of forward dimuons in Pb-Pb UPC.}
\end{minipage} \hfill
\begin{minipage}{0.55\linewidth} \begin{center} \vspace*{-0.5cm} \hspace*{-0.cm}
\mbox{\includegraphics[width=1.1\linewidth]{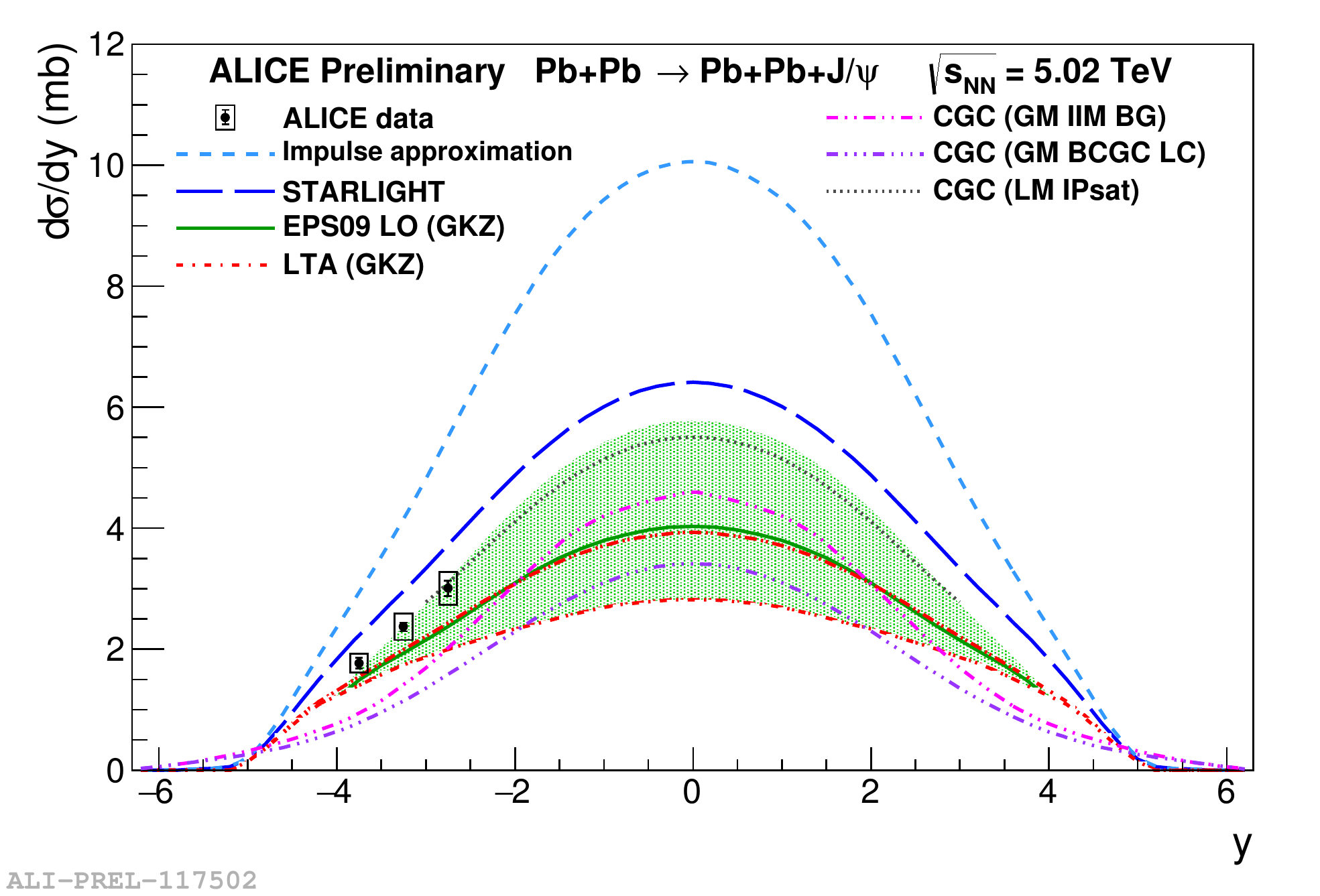}} \end{center} \vspace*{-0.4cm}
\caption{\label{jpsiresult} Forward \jpsi photoproduction cross section in Pb-Pb UPC at $\sqrt{s_{\rm NN}}=5.02$ TeV.}
\end{minipage} \hfill
\end{figure}

 The selected data sample contains around nine thousand events. The dimuon invariant mass spectrum is shown in Fig.~\ref{jpsisignal} together with a fit to a
Crystall Ball function for the \jpsi resonance and a polynomial function for the $\gamma\gamma \rightarrow \mu\mu$ continuum.
According to the fit result, around six thousand events with \jpsi produced in UPC are reconstructed.

The transverse momentum distribution of dimuons with the invariant mass in the range 2.85 $<$ M$_{\mu\mu} <$ 3.35 GeV/{\it c}$^2$
is then fitted with Monte-Carlo templates corresponding to coherent \jpsi photoproduction and different sources of the 
background for the present measurement - incoherent \jpsi photoproduction, dimuon production in \gaga collisions 
and feed-down from $\psi$(2S) decays as well. The background subtracted data were corrected for the detector acceptance 
with a simulation made by the STARlight event generator \cite{sl}. 
\par The preliminar ALICE result on the coherent \jpsi photoproduction cross section 
in Pb-Pb UPC is shown in Fig.~\ref{jpsiresult} together with theoretical calculations based on
the impulse approximation (no nuclear effects), STARlight (vector meson dominance model), predictions
using the Color Glass Condensate approach \cite{cgc} and calculations based on the EPS09 framework and on the Leading Twist
Approximation (LTA) \cite{gkz}. The data support moderate gluon shadowing in nuclei and agree with calculations incorporating shadowing according to EPS09.

\section{Coherent \rrho photoproduction}
The \rrho vector mesons provide a sizable contribution to the hadronic structure of the photon. The total
$\gamma p$ cross section contains (depending on energy) (10-20)\% of the $\gamma+p\rightarrow\rho^0+p$ contribution.
The \rrho photoproduction off a nuclear target is usually modeled with the
Glauber approach coupled to the Vector Meson Dominance model.
The large value of the cross section means that for heavy nuclei the target appears like a black disk and 
the total photonuclear cross section reaches its saturation. Thus the study of coherent \rrho photoproduction in UPC at the LHC
is aimed at investigating nuclear shadowing effects in the nonperturbative regime.
\par
\begin{figure}[!h]
\vfill \begin{minipage}{0.4\linewidth} \begin{center} \vspace*{-0.1cm} \hspace*{-1.cm}
\mbox{\includegraphics[width=1.1\linewidth]{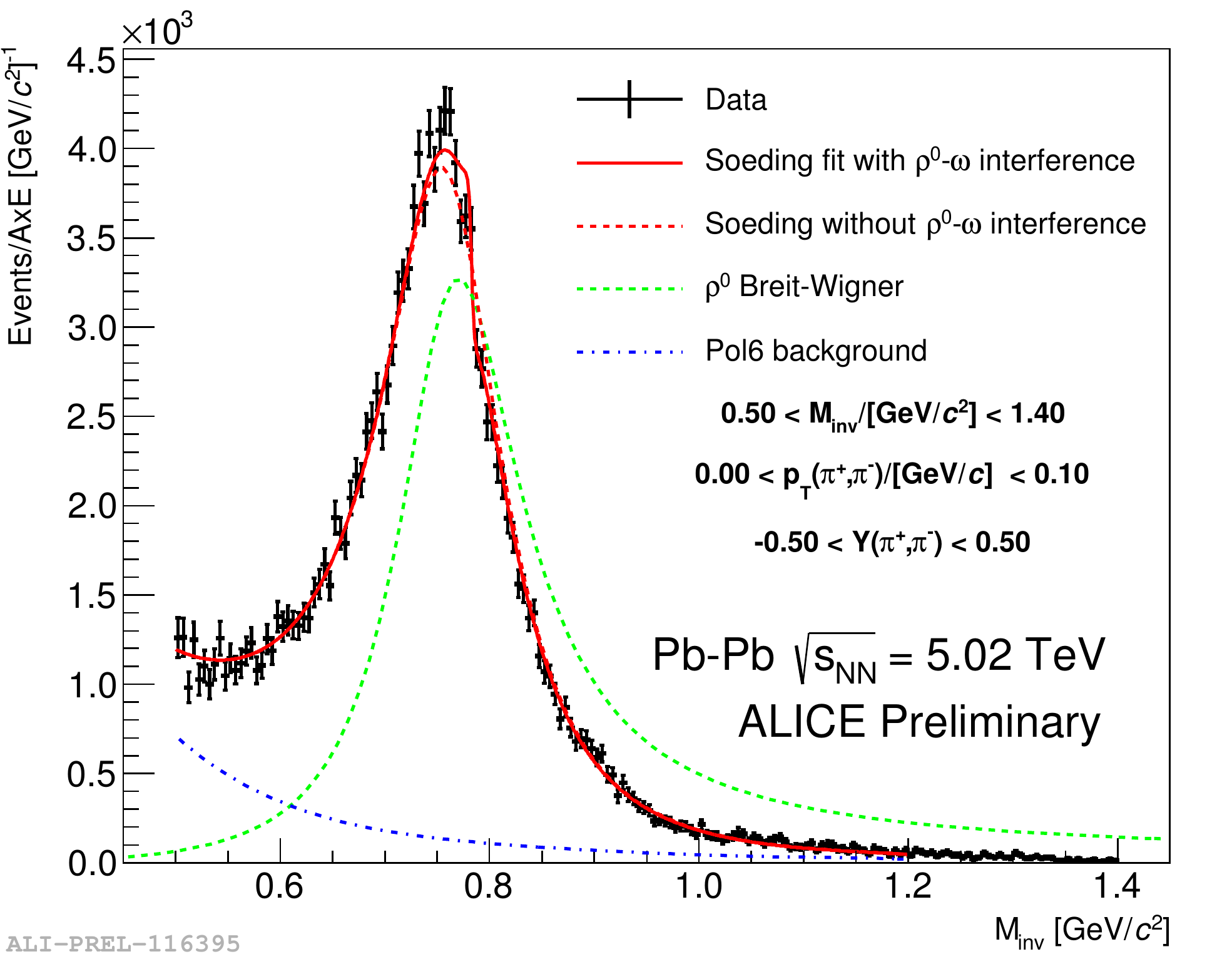}} \end{center} \vspace*{-0.cm}
\caption{\label{rhosignal} Fit to the dipion mass spectrum with a S$\ddot{\rm o}$ding model with and without 
including $\rho^0-\omega$ interference.}
\end{minipage} \hfill
\begin{minipage}{0.5\linewidth} \begin{center} \vspace*{-0.1cm} \hspace*{-0.cm}
\mbox{\includegraphics[width=1.1\linewidth]{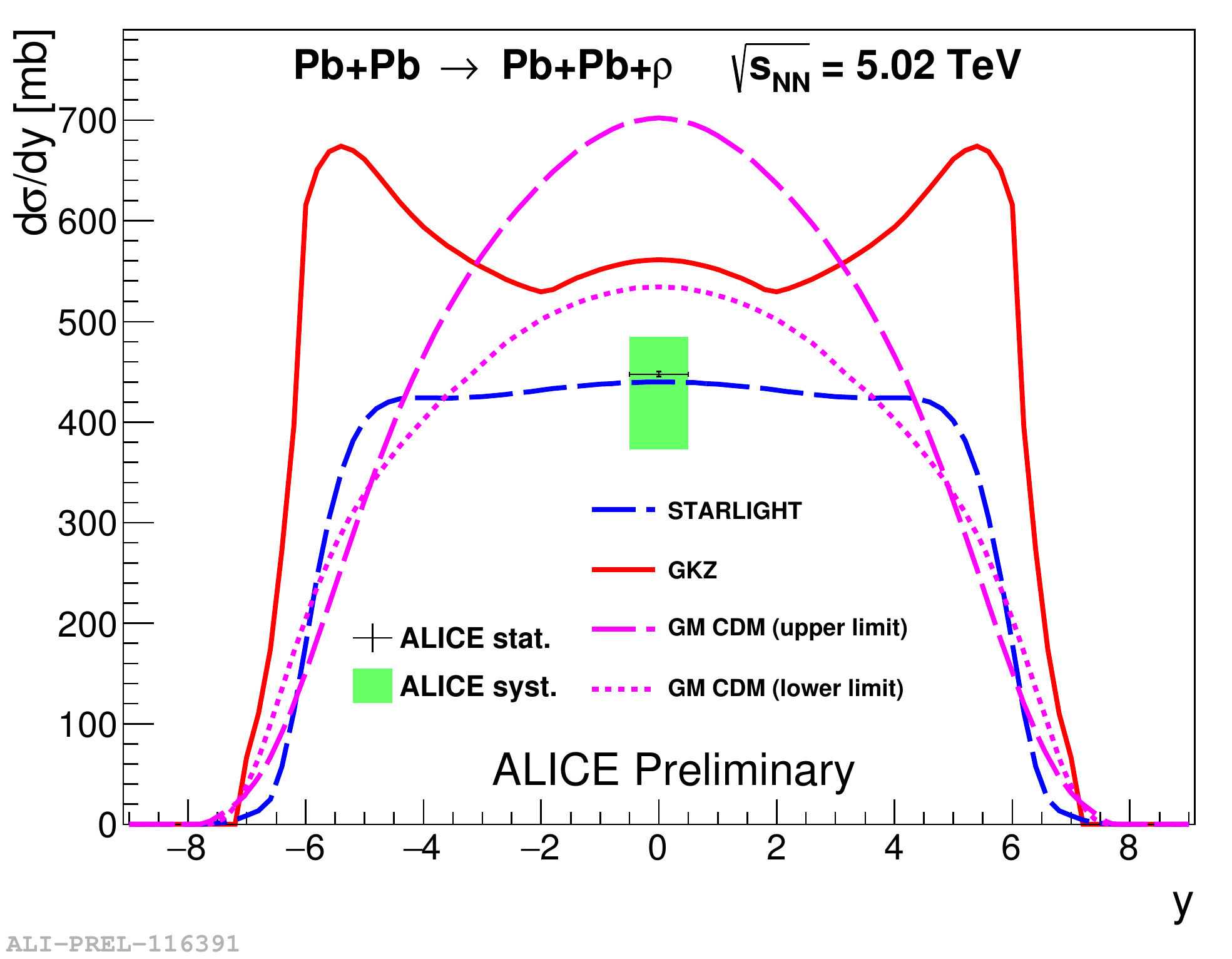}} \end{center} \vspace*{-0.4cm}
\caption{\label{rhoresult} Mid-rapidity \rrho photoproduction cross section in Pb-Pb UPC at $\sqrt{s_{\rm NN}}=5.02$ TeV.}
\end{minipage} \hfill
\end{figure}

 The results on \rrho mid-rapidity photoproduction in Pb-Pb UPC at $\sqrt{s_{\rm NN}}=5.02$ TeV are presented here.
The data were taken with a dedicated trigger involving vetoes on any activity in forward scintillators 
and transverse back-to-back topology of SPD hits. The \rrho signal was observed in the
$\pi^+\pi^-$ channel in the rapidity range $|$y$| <$ 0.5. The pion-pair mass spectrum, corrected for experimental effects 
was fitted with a S$\ddot{\rm o}$ding function \cite{sod} as it is shown in Fig.~\ref{rhosignal}. 
The background coming from dimuon pairs produced in two-photon interactions was estimated with STARlight, while
a contribution of incoherently produced \rrho mesons - by the fitting of the pair p$_T$ spectrum.
\par The preliminar ALICE result on coherent \rrho photoproduction in Pb-Pb UPC at
$\sqrt{s_{\rm NN}}=5.02$ TeV is shown in Fig.~\ref{rhoresult} in comparison to model predictions. 
The model based on the Glauber approach and photon inelastic diffraction into large masses \cite{gkz} leads
to essential nuclear shadowing effects but still overpredicts the data. 
The measurement appears
to be in agreement with the STARlight model, which neglects the elastic part of the total $\rho^0$N cross section. The model
based on the color-dipole approach and the Color Glass Condensate formalism \cite{gm} is also found to be above the data.

\end{document}